\documentclass[3p,times,procedia]{elsarticle}
\usepackage{nupha_ecrc}

\volume{00}

\firstpage{1}

\journalname{Nuclear Physics A}

\runauth{}

\jid{nupha}

\jnltitlelogo{Nuclear Physics A}

\usepackage{amssymb}

\usepackage[figuresright]{rotating}

\usepackage{amsmath,bm}

\newcommand{\be}{\begin{equation}}
\newcommand{\ee}{\end{equation}}
\newcommand{\ba}{\begin{eqnarray}}
\newcommand{\ea}{\end{eqnarray}}
\newcommand{\ban}{\begin{eqnarray*}}
\newcommand{\ean}{\end{eqnarray*}}

\def\v2{\mbox{$v_2$}}

\begin{document}

\begin{frontmatter}



\dochead{XXVIIIth International Conference on Ultrarelativistic Nucleus-Nucleus Collisions\\ (Quark Matter 2019)}

\title{Magnetic Field in the Charged Subatomic Swirl}


\author[1]{Xingyu Guo\corref{corl}}
\ead{guoxy@m.scnu.edu.cn}
\author[2]{Jinfeng Liao}
\author[1]{Enke Wang}
\address[1]{Guangdong Provincial Key Laboratory of Nuclear Science, Institute of Quantum Matter, South China Normal University, Guangzhou 510006, China}
\address[2]{Physics Department and Center for Exploration of Energy and Matter, Indiana University, 2401 N Milo B. Sampson Lane, Bloomington, IN 47408, USA.}

\begin{abstract}
We report a novel relation between rotation and magnetic field in a charged fluid system: there is naturally a magnetic field along the direction of fluid vorticity due to the currents associated with the swirling charges.  This general connection is demonstrated using a fluid vortex. Applying the idea to heavy ion collisions we propose a new mechanism for generating in-medium magnetic field with a relatively long lifetime. We estimate the magnitude of this new magnetic field  in the Au-Au colliding systems across a wide span of collisional beam energy. Such a magnetic field is found to increase rapidly  toward lower beam energy and could account for a significant amount of the experimentally observed global polarization difference between hyperons and anti-hyperons.
\end{abstract}


\end{frontmatter}


\section{Introduction}

There have been increasing interests in the properties of many-body systems in the existence of strong magnetic field or rotation. These interests come across a wide range of fields of physics, including condensed matter physics, cold atomic gases, astrophysics and nuclear physics, see e.g. recent reviews in~\cite{Bzdak:2019pkr,Fukushima:2018grm,Kharzeev:2015znc,Fetter:2009zz}.  These extreme fields can induce nontrivial anomalous chiral transport effects such as the Chiral Magnetic Effect (CME)~\cite{Kharzeev:2007jp,Fukushima:2008xe,STAR_LPV1,STAR_LPV_BES} and Chiral Vortical Effect (CVE)~\cite{Son:2009tf,Kharzeev:2010gr,Landsteiner:2011iq}. They can also change the phase structures and influence novel phenomena like fermion pairings~\cite{Jiang:2016wvv}. In this contribution we focus on a new state of matter with extremely high temperature known as a quark-gluon plasma (QGP) that is created in relativistic heavy ion collisions. There exist the strongest magnetic fields as well as the largest fluid vorticity in such collisions. There have been enthusiastic efforts searching for novel effects associated with these extreme fields, see e.g. the latest status discussed in ~\cite{Bzdak:2019pkr}.

The magnetic field in heavy ion collisions plays a central role for the CME signal in QGP. While the initial vacuum magnetic field from spectators reaches a few times pion-mass-square, its lifetime is too short compared to the formation of QGP~\cite{Bloczynski:2012en,McLerran:2013hla,Tuchin:2015oka,Inghirami:2016iru,Gursoy:2018yai}. 
The fluid vorticity originates from the large angular momentum carried by the colliding system and has been quantitatively simulated with various tools~\cite{Becattini:2013vja,Csernai:2013bqa,Becattini:2015ska,Jiang:2016woz,Shi:2017wpk,Deng:2016gyh,Becattini:2016gvu,Li:2017slc,Sun:2017xhx}. The observable effects of such vorticity include global spin polarization of produced hadrons~\cite{Liang:2004ph,Gao:2007bc,Voloshin:2004ha,Betz:2007kg,Becattini:2007sr}.   Recently the STAR Collaboration at RHIC measured this effect for the $Lambda$ and anti $\Lambda$ hyperons~\cite{STAR:2017ckg}, showing an average fluid vorticity of about $10^{21}\, s^{-1}$. There is also a non-zero difference between the polarization of hyperons and anti-hyperons. Attempts were made to explain this puzzle but so far inconclusive~\cite{Becattini:2016gvu,Csernai:2018yok,Muller:2018ibh}. The splitting could arise from  a sufficiently long-lived magnetic field  but it is unclear how this kind of field exists.

In this contribution (based on publication~\cite{Guo:2019mgh}), we report a possible mechanism based on a novel relation between rotation and magnetic field in a charged fluid system that may provide resolutions to both puzzles discussed above. The basic idea is that a magnetic field naturally arises due to the currents associated with the swirling charges. We will demonstrate this general connection and then apply it to heavy ion collisions as a new mechanism for generating long-lived in-medium magnetic field. We further show that the so-obtained magnetic field could induce a significant splitting between the hyperon and anti-hyperon global polarization.

\section{Magnetic Field of  A Charged Fluid Vortex}

Let us use the simplest case, a classical relativistic charged particle (with charge $q e$ and mass $m$) undergoing uniform circular motion at an angular speed $\omega$, to demonstrate the mechanism. In this case, one can easily show the magnetic field from this swirling charge $\mathbf{B} \propto (qe) \bm{\omega}$. 

We next consider a cylinder-like fluid vortex of a transverse size $R_0$, with nonzero average vorticity $\bm{\bar{\omega}}$  and nonzero charge density $n$. By solving the Maxwell's equations in this system, one can derive a key result that connects the magnetic field with the vorticity: 
\begin{eqnarray} \label{eq_omega_B}
e\bm{\bar{B}} = \frac{e^2}{4\pi} \  n \ (\pi R_0^2) \ \bm{\bar{\omega}} =  \frac{e^2}{4\pi} \  n \ A \ \bm{\bar{\omega}}
\end{eqnarray} 
where $A=\pi R_0^2$ is the transverse area of the fluid vortex. The detailed derivation can be found in~\cite{Guo:2019mgh}.  
This relation suggests that there always exists a magnetic field in a charged fluid vortex, whose average value is linearly proportional to the charge density as well as the average fluid vorticity.  This simple relation suggests a possible new mechanism for generating magnetic field in heavy ion collisions.

\section{Magnetic Field in Heavy Ion Collisions}

In heavy ion collisions, there are vorticity structures and a nonzero net charge density in the created QGP. Given the connection between magnetic field and the vorticity in a charged fluid in Eq.\eqref{eq_omega_B},  we propose this as a new mechanism for the generation of long-lived magnetic field.  
In the following we will estimate the magnitude of this new magnetic field and examine the implication for relevant experimental observables. 

\begin{figure*}[htb!]
\includegraphics[width=16cm]{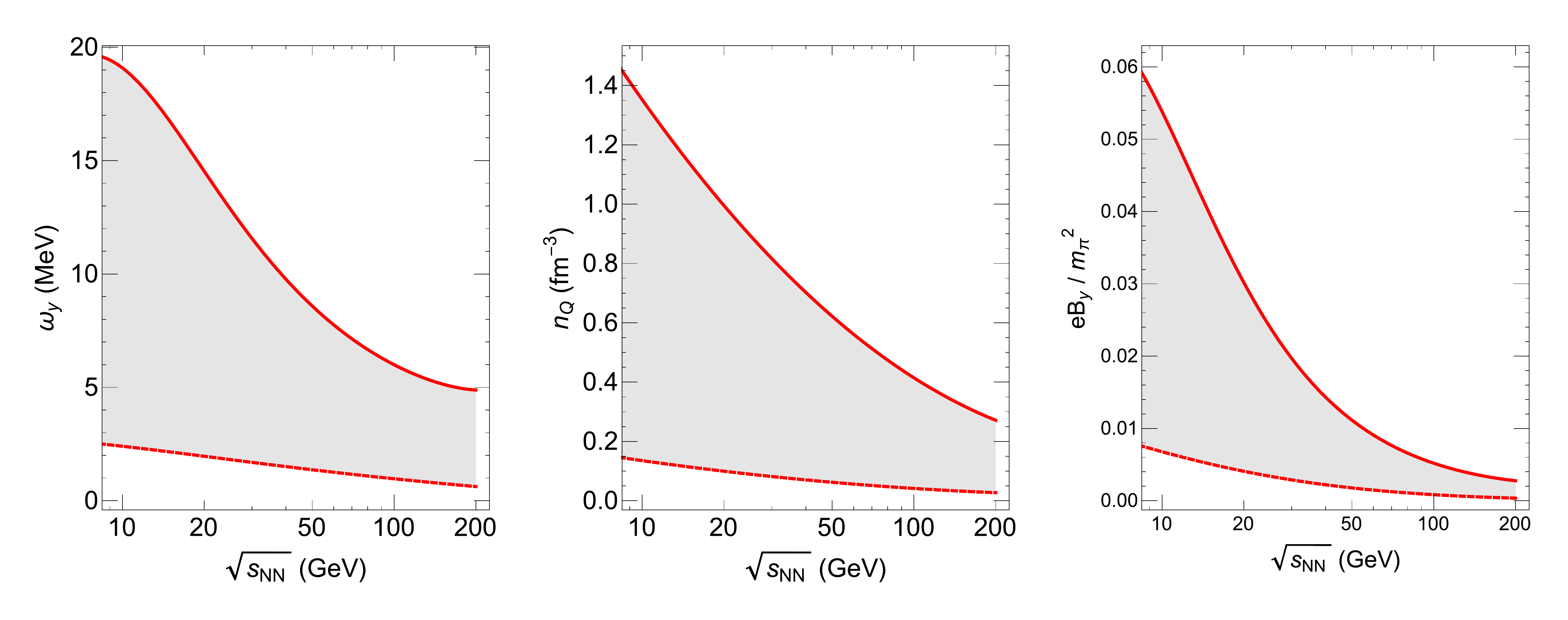}
\caption{\label{fig1} (color online) The vorticity $\omega_y$ (left), charge density $n_Q$ (middle) and magnetic field $e\bar{B}$ (right) as functions of collisional beam energy $\sqrt{s_{NN}}$, with solid/dashed curves in each panel representing an upper/lower estimates.}
\end{figure*}

One can extract average vorticity $\omega_y$ (along the out-of-plane direction) and charge density $n$ for a wide range of beam energy from AMPT simulations~\cite{Jiang:2016woz,Shi:2017wpk,Li:2017slc}. We take $(20-50)\%$ centrality of AuAu collisions at RHIC in the $(10\sim 200)\rm GeV$ energy region which is relevant to STAR measurements in~\cite{STAR:2017ckg}. The average vorticity and charge density both decrease in time as the fireball expands. We show in Fig.~\ref{fig1} such average vorticity (left) and charge density (middle) values as a function of beam energy $\sqrt{s}$ for an early time moment $\tau=0.5\rm fm$ (solid curve) and a late time moment $\tau=5\rm fm$ (dashed curve), with the shaded band indicating the expected range. Towards lower beam energy  both vorticity and charge density increase quickly. For the transverse area of fluid vortex, we can use  $A\sim \pi R_0^2 $ with $R_0\sim 4\rm fm$ as an order-of-magnitude estimate.   
Putting all these together into Eq.\eqref{eq_omega_B}, we now have an estimate for the magnetic field $e \bar{B} $ from the charged fluid vortex, as shown in Fig.~\ref{fig1} (right). As one can see, a magnetic field with the order of magnitude $\sim 0.01 m_\pi^2$ could be generated. This magnetic field increases strongly toward lower beam energy. When compared with the initial magnetic field from spectators, its peak value is smaller, but its lifetime is much longer and may make considerable contributions to interesting effects induced by magnetic field.

\begin{figure}[htb!]
\begin{center}
\includegraphics[width=8.8cm]{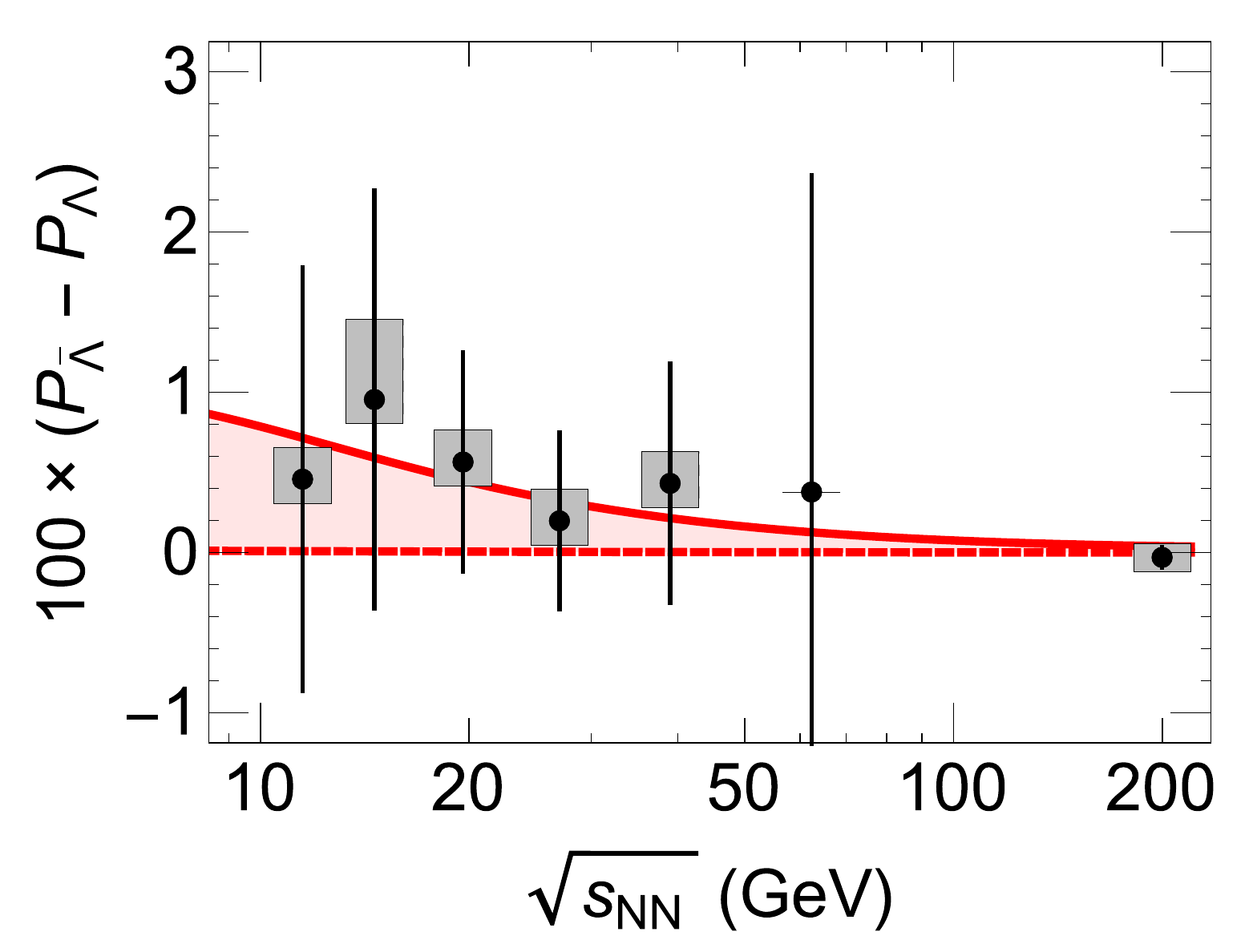}
\caption{\label{fig2} (color online) The induced polarization difference between hyperons and anti-hyperons, $\Delta P = P_{\bar{\Lambda}} - P_\Lambda$ as a function of collisional beam energy $\sqrt{s_{NN}}$, in comparison with STAR data~\cite{STAR:2017ckg}.   The solid/dashed curves are obtained from the upper/lower estimates for $e\bar{B}$  (see solid/dashed curves respectively in Fig.~\ref{fig1} right panel).}\vspace{-0.25in}
\end{center}
\end{figure}

Such a magnetic field would contribute  to the splitting of polarization of hyperons and anti-hyperons due to their opposite magnetic moments~\cite{Patrignani:2016xqp}. With the presence of a magnetic field at freeze-out, one expects: 
\begin{eqnarray}  
\Delta P \equiv P_{\bar\Lambda} - P_{\Lambda} \simeq \frac{2|\mu_\Lambda | \bar{B}}{T_{fo}} 
\end{eqnarray}
where $|\mu_\Lambda|  = 0.613 \mu_N=\frac{0.613 \ e}{2 M_N}$ with $M_N=938\rm MeV$~\cite{Patrignani:2016xqp} and $T_{fo} = 155\rm MeV$. The induced polarization difference $\Delta P$ as a function of beam energy is shown in Fig.~\ref{fig2}, in comparison with STAR data. Again the solid/dashed curves correspond to the upper/lower estimates for 
$e\bar{B}$, with the shaded band between them giving the expected range. Although current experimental error bars are still large, our proposed new mechanism can induce a considerable difference in the hyperon/anti-hyperon polarizations that is of the same order of magnitude as the experimental measurements. This mechanism also has a dependence of collisional beam energy in consistency with data.  
Its semi-quantitative success  motivates a more quantitative future computation using state-of-art modelings~\cite{Jiang:2016wve,Shi:2017cpu,Guo:2017jxs}.

\section{Summary}

We have suggested a novel link between rotation and magnetic field in a charged fluid system. Based on this idea we have derived a concrete result for magnetic field generated by a charged fluid vortex. We have further applied this idea to heavy ion collisions as a new mechanism for generating long-lived in-medium magnetic field. Estimates have been made for  this new magnetic field across a wide range of collision beam energy, with a rapidly increasing trend towards lower beam energy. Finally we have demonstrated how it could provide a possible explanation for the experimentally observed puzzle about polarization difference between hyperons and anti-hyperons.   \\

{\em Acknowledgements.---}The authors thank Shuzhe Shi for very helpful discussions. This work is supported in part by the NSFC Grants No. 11435004 and No. 11735007, by the NSF Grant No. PHY-1913729 and by the U.S. Department of Energy, Office of Science, Office of Nuclear Physics, within the framework of the Beam Energy Scan Theory (BEST) Topical Collaboration.











{}

\end{document}